\documentclass[aps,twocolumn,showpacs,superscriptaddress,preprintnumbers,amsmath,amssymb]{revtex4}
\usepackage{graphicx}
\usepackage{dcolumn}
\usepackage{bm}
\usepackage{color}
\usepackage{adjustbox}
\usepackage{multirow}
\usepackage{siunitx}

\begin{document}

\title{
\large Experimental Observation of Non-Exponential Auger-Meitner Decay of Inner-Shell-Excited CO}

\author{M.~Weller}
\author{G.~Kastirke}
\author{J.~Rist}
\affiliation{Institut f\"{u}r Kernphysik, Goethe-Universit\"{a}t Frankfurt, Max-von-Laue-Stra{\ss}e 1, 60438 Frankfurt am Main, Germany}
\author{C.~Goy}
\affiliation{Deutsches Elektronen-Synchrotron DESY, Notkestra{\ss}e 85, 22607 Hamburg, Germany}
\author{A.~Khan}
\affiliation{Indian Institute of Science Education and Research, Bhopal, Madhya Pradesh 462066, India}
\author{M.~Kircher}
\author{C.~Rauch}
\author{L.~Ph.~H.~Schmidt}
\affiliation{Institut f\"{u}r Kernphysik, Goethe-Universit\"{a}t Frankfurt, Max-von-Laue-Stra{\ss}e 1, 60438 Frankfurt am Main, Germany}
\author{N.~Sisourat}
\affiliation{Sorbonne Universit\'{e}, CNRS, Laboratoire de Chimie Physique Mati\`{e}re et Rayonnement, UMR 7614, F-75005 Paris, France}
\author{M.~S.~Schöffler}
\author{R.~Dörner}
\affiliation{Institut f\"{u}r Kernphysik, Goethe-Universit\"{a}t Frankfurt, Max-von-Laue-Stra{\ss}e 1, 60438 Frankfurt am Main, Germany}
\author{F.~Trinter}\email{trinter@fhi-berlin.mpg.de}
\affiliation{Molecular Physics, Fritz-Haber-Institut der Max-Planck-Gesellschaft, Faradayweg 4-6, 14195 Berlin, Germany}
\author{T.~Jahnke}\email{till.jahnke@xfel.eu}
\affiliation{European XFEL, Holzkoppel 4, 22869 Schenefeld, Germany}
\affiliation{Max-Planck-Institut f\"{u}r Kernphysik, Saupfercheckweg 1, 69117 Heidelberg, Germany}

\date{\today}

\begin{abstract}
Electronically excited atoms or molecules may deexcite by emission of a secondary electron through an Auger-Meitner decay. This deexcitation process is typically considered to be exponential in time. This is strictly speaking, however, only true for the case of an atom. Here, we present a study experimentally demonstrating the non-exponential time dependence of the decay of an inner-shell hole in a diatomic molecule. In addition, we provide an intuitive explanation for the origin of the observed variation of the molecular lifetimes and their dependence on the kinetic energy of the ionic fragments measured in coincidence with the photoelectrons.
\end{abstract}


\maketitle

Electronically excited atoms or molecules may undergo a transition into an energetically favored state by emission of a photon or a secondary electron. The latter process is known since more than hundred years as the Auger decay (or more recently Auger-Meitner decay, as Lise Meitner was the first to observe this process) \cite{Meitner1922, Auger1925}. As most decay processes, the Auger-Meitner decay is of exponential nature, i.e., an ensemble of excited species will deexcite exponentially over time. This behavior is exploited, for example, in electron spectroscopy as it yields a Lorentzian line shape in the energy spectra of the excited state, and the width of that Lorentzian distribution is directly connected to the lifetime of the decaying state. Strictly speaking, however, this exponential decay behavior is only occurring if longer time scales are considered \cite{Giacosa2014} and in case of an \textit{atomic} Auger-Meitner decay. Neglecting the former, i.e., on macroscopic time scales, molecules are expected to exhibit in most cases an intrinsically non-exponential decay behavior because of the coupling of electronic and nuclear degrees of freedom. The decay probability is related to the (spatial) overlap of the states involved in the decay, and this overlap may differ strongly depending on the exact internuclear distances of the atoms of the molecule. Hence, the decay probability is expected to vary, for example, as the internuclear distances vary due to molecular vibrations \cite{Kaspar1979}. This theoretically predicted (and very general) molecular effect is, however, mostly neglected in the interpretation of molecular spectra, and the corresponding theoretical modeling is performed within the \textit{constant resonance width approximation} \cite{Correia1984}. This is sufficient in most cases, as the corresponding molecular electron spectra are typically very complex and the effect caused by the lifetime being not constant over the nuclear coordinates on these spectra is usually very small. More detailed studies, which take into account the molecular ion generated in the electronic decay (and/or by the initial electronic excitation) \cite{Ulrich2008} have, however, the potential to unravel this property of the Auger-Meitner decay. In this Letter, we present a corresponding experimental study on the K-shell ionization of carbon monoxide molecules demonstrating directly the non-exponential behavior of the molecular Auger-Meitner decay occurring subsequent to the ionization.

\begin{figure}[htbp]
    \includegraphics[width=1.0\linewidth]{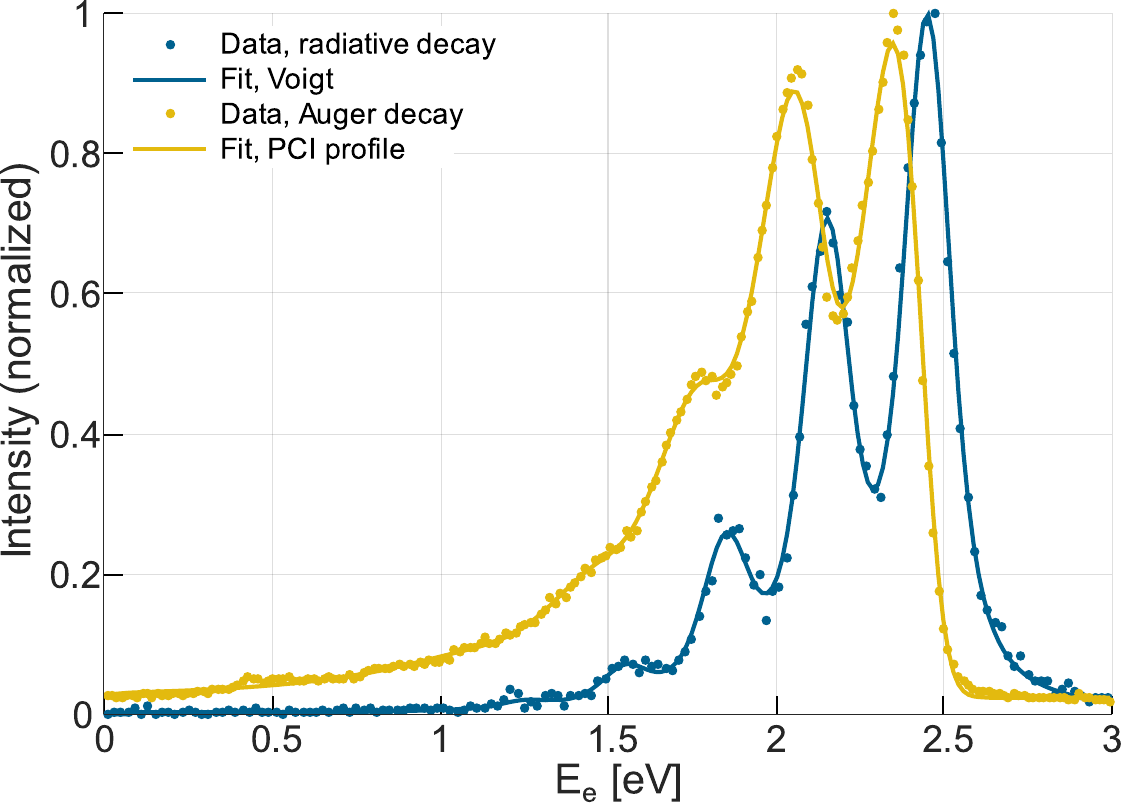}
    \caption{\label{fig:comp_radAuger} CO C~1s photoelectron spectrum postselected for subsequent radiative (blue) and Auger-Meitner decay (yellow) recorded simultaneously with an experimental resolution of $\approx$80~meV. The spectrum obtained for the Auger-Meitner-decaying molecule shows typical PCI-induced distortions, i.e., non-symmetric broadened lines skewed to smaller energies. Fitted to the data are Voigt profiles for the radiative-decay case and PCI profiles convoluted with a Gaussian function for the Auger-Meitner-decaying channel.}
\end{figure}

We employed cold target recoil ion momentum spectroscopy (COLTRIMS) \cite{Doerner2000, Ullrich2003, Jahnke_JESRP2004} for the measurements presented in this Letter. The synchrotron light ($h\nu = 297.3$~eV) for ionizing the CO molecules was provided by beamline U49-2\_PGM-1 \cite{Kachel2016} at the BESSY~II electron storage ring operated by the Helmholtz-Zentrum Berlin für Materialien und Energie running in single-bunch mode. The photon bandwidth was $\Delta$E~$<$~31~meV. The synchrotron beam was crossed at right angle with a supersonic gas jet consisting of CO molecules in the vibronic ground state. Ions and electrons generated in the photoreaction (and the subsequent Auger-Meitner decay) were guided by weak electric fields to two time- and position-sensitive microchannel-plate detectors with hexagonal delay-line position readout \cite{jagutzki02nim}. The detectors had an active area of 120~mm diameter. From the positions-of-impact and the times-of-flight, the initial vector momentum of each detected particle was inferred by reconstructing the particle's trajectory inside the COLTRIMS analyzer in an offline analysis. In order to achieve a sufficiently high photoelectron-energy resolution, we employed in addition to a time-focusing geometry an electrostatic lens in the electron arm of the spectrometer \cite{Lebech2002}. This lens was used to minimize the effect of the finite size of the interaction region of the gas jet and the light beam. The overall length of the electron arm was 397~mm. The ion arm consisted of a homogeneous acceleration field with a strength of E$_{acc} = 7$~V/cm and had a total length of 31~mm. Electrons up to a kinetic energy of approximately 3~eV and molecular breakups into C$^+$ and O$^+$ up to a kinetic energy release of KER~=~25~eV were detected with full solid-angle coverage.

Figure~\ref{fig:comp_radAuger} shows the measured photoelectron spectrum after carbon K-shell ionization of the molecule. The blue distribution depicts the kinetic energy of the photoelectron for events where the core-ionized molecule decays radiatively yielding a CO$^+$ molecular ion in the final state. The peaks observable here belong to the different vibrational states {($\nu' = 0...4$)} of the core-ionized CO$^+$ cation. The experimental data points have been fitted by using a sum of Voigt profiles consisting of a Lorentzian function convoluted with a Gaussian function accounting for the finite energy resolution of our experimental apparatus (80~meV for the electron-energy resolution) and the employed photons (31~meV for the measured photon-energy bandwidth).
The yellow data points belong to events in which after the K-shell ionization an Auger-Meitner decay occurred. This decay typically leads to a breakup of the molecule into C$^+$ and O$^+$, and we can discriminate these cases in our recorded data set by measuring the two generated fragment ions in coincidence with the photoelectron. Due to post-collision interaction (PCI) of the Auger electron and the photoelectron, the line shape of the individual photoelectron peaks becomes asymmetric (as the photoelectron is decelerated, while the Auger electron is accelerated \cite{Niehaus1977}). The amount of deceleration depends on the initial kinetic energy of the interacting electrons and on the decay time (i.e., the emission time of the Auger electron relative to the photoelectron ejection) \cite{Schuette2012, Guillemin2012}. We will use the latter dependence to trace down the non-exponential decay features in the following. The yellow line is a fit of a quantum mechanical PCI model by Armen \textit{et al.} \cite{Armen:1987} convoluted with a Gaussian function accounting for experimental uncertainties to the spectra. Fit parameters are the intensity factors $a_n$ of the different vibrational states ($\nu' = 0...4$), decay times $\tau_n$ of these states, and an addend for a constant measurement background. Please see below for details on the parameters used for the fitting. This fitting procedure implies an exponential decay of the K-shell vacancy (despite different lifetimes were assumed for the different vibrational levels of the cation). This result gives an indication how hard it is to trace non-exponential decay properties using classical electron spectroscopy.

\begin{figure}[htbp]
    \includegraphics[width=\linewidth]{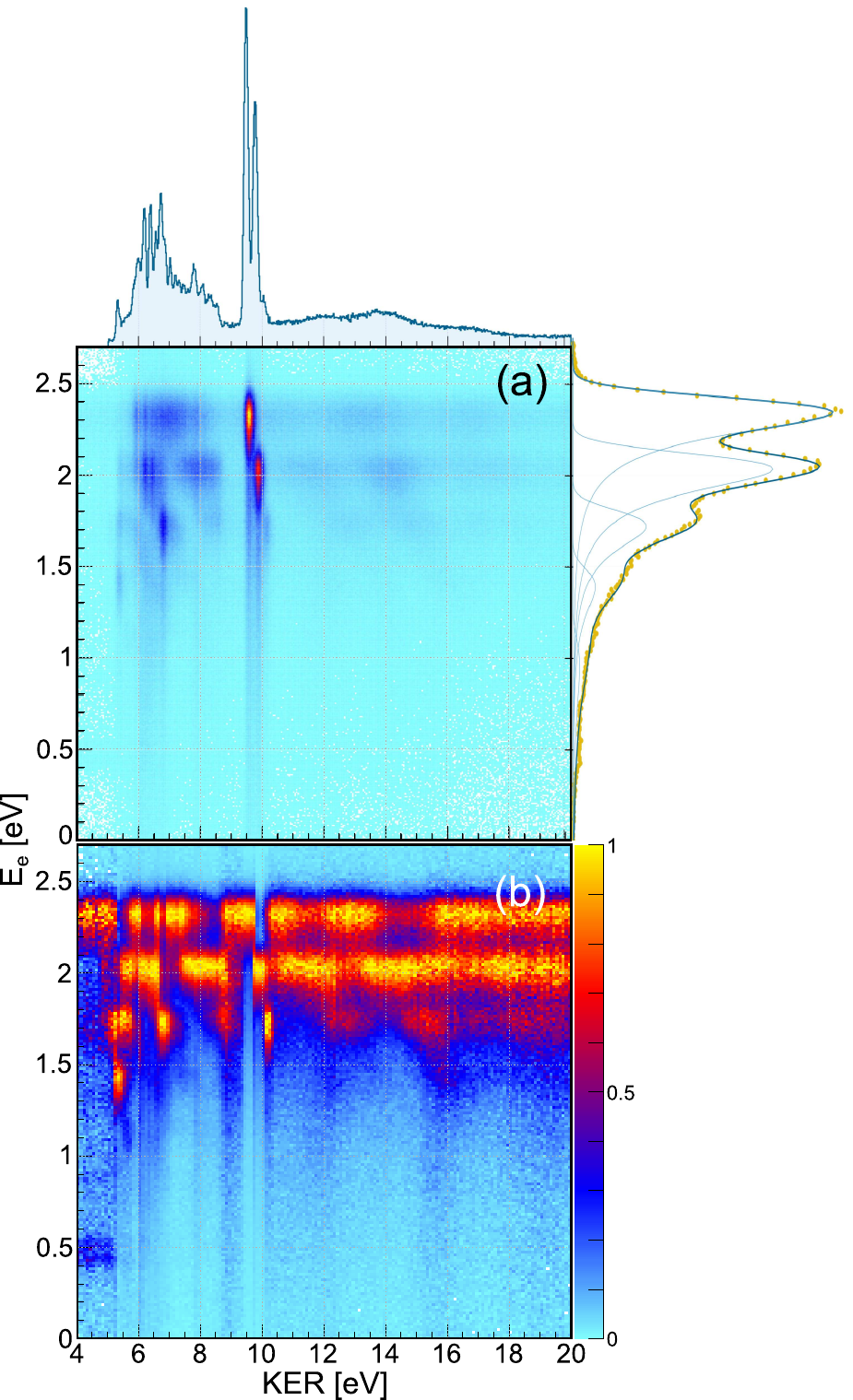}
    \caption{\label{fig:KER_Ee_histo}Energy correlation between photoelectron and molecular-dication kinetic energy release. Horizontal axis: KER of the C$^+$ and O$^+$ fragments, vertical axis: photoelectron energy. Panel~(a): counts as measured, panel~(b): each column is normalized to the maximum counts in the respective column.}
\end{figure}

\begin{figure}[htbp]
    \includegraphics[width=1.0\linewidth]{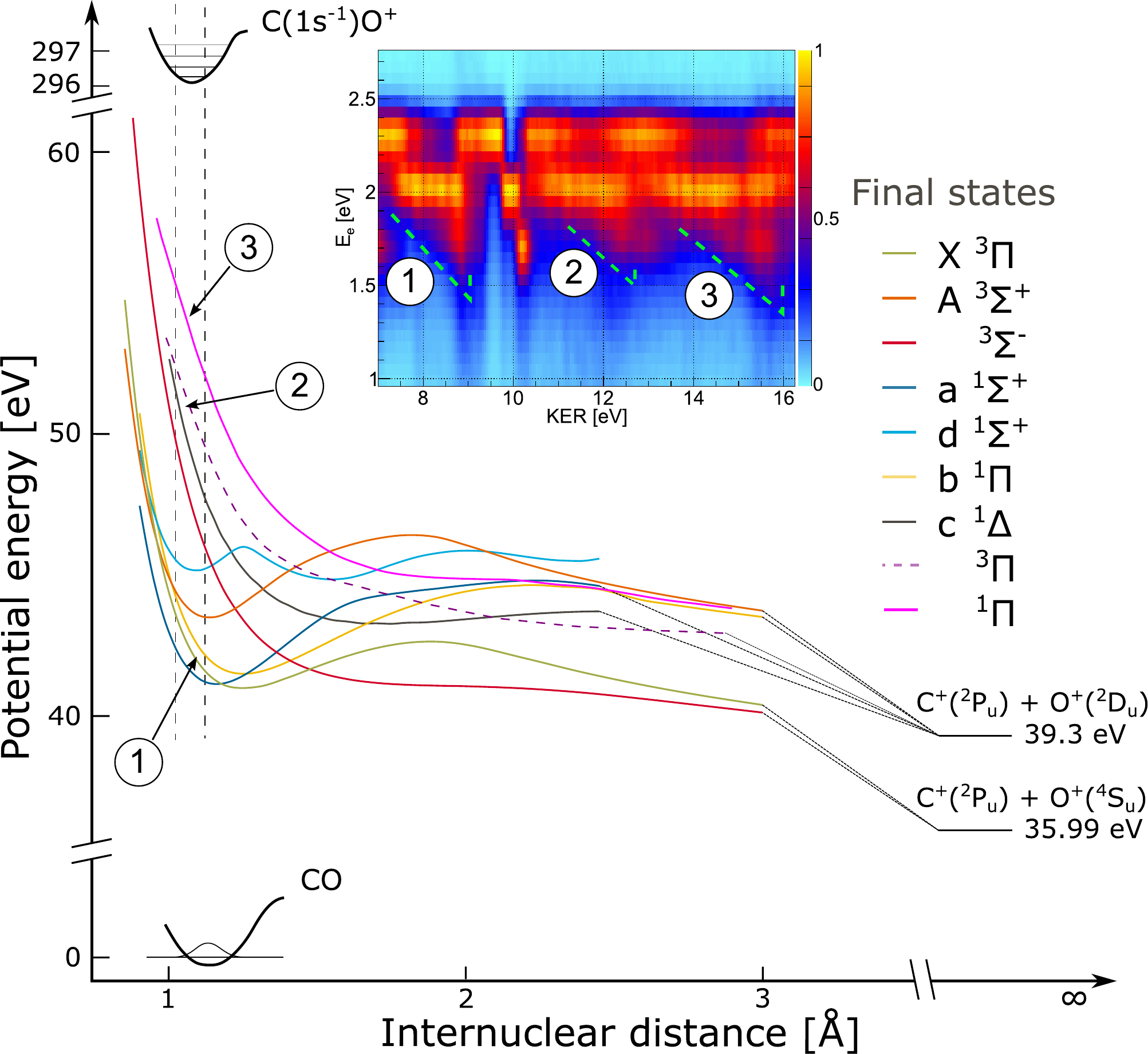}
    \caption{\label{fig:Pot_curve}Potential energy curves of the states involved in the Auger-Meitner decay after K-shell photoionization of CO. The data points of the curves of the neutral ground state and the final state were taken from Ref.~\cite{Eland2004}, the curves of the initial states including the vibrational levels stem from Ref.~\cite{Kempgens1997}. The parallel vertical lines mark the Franck-Condon regions of the Auger-Meitner decays of the C~1s$^{-1}$ states of CO$^+$. Their positions were taken from Ref.~\cite{Puettner2007}. The dissociative limits of the C$^+$~+~O$^+$ states stem from Ref.~\cite{Lablanquie1989}.}
\end{figure}

Figure~\ref{fig:KER_Ee_histo} provides a more detailed view on the electronic decay of the ionized carbon K shell. Panel~(a) depicts the kinetic energy release (KER), i.e., the sum of the kinetic energies of the C$^+$ and O$^+$ fragment ions in their center-of-mass frame, versus the photoelectron kinetic energy. Projections of the coincidence map are depicted at the top and at the right showing the integrated ion and electron spectra. The electron distribution on the right is identical to the yellow one in Fig.~\ref{fig:comp_radAuger}. The coincidence map provides detailed information on the states involved in the Auger-Meitner decay. The relevant potential energy curves of the CO$^+$ and CO$^{++}$ ions are shown in Fig.~\ref{fig:Pot_curve}, indicating the manifold of different decay channels, which may contribute to the Auger-Meitner process. For example, the most prominent set of peaks in the KER distribution located approximately at KER~=~10~eV is generated by a transition of the cation within the Franck-Condon region (marked by the vertical dashed lines) to vibrational states of the d$^1\Sigma^+$ curve, which then couple onto the $^3\Sigma^-$ curve, leading to the ground state of the fragmented molecular dication, i.e., the C$^+$($^2$P)~+~O$^+$($^4$S) final state \cite{Cederbaum1991, Lundqvist1995, Schimmelpfennig1996, Weber2003}. The coincidence map shows nicely that different sub-peaks of this set are connected to the different vibrational states of the cation \cite{Lundqvist1995}.

While the spectroscopic assignment is of interest on its own, it is not the scope of this Letter. Instead, we generated Fig.~\ref{fig:KER_Ee_histo}(b) by normalizing the maximum value of each column of the coincidence map shown in panel~(a) to 1. This way, the dominating effect of the shape of the KER distribution is removed from the representation. The normalized histogram shows that depending on the KER the amount of low-energy electrons differs quite strongly. For example, in a range around 8.0~eV~$<$~KER~$<$~8.5~eV the amount of electrons with an energy ${E_e}_\gamma < 1$~eV is much less as compared to the adjacent region at a KER of approximately 9~eV. In this low-electron-energy region, we do not expect contributions from vibrational states of the cation, as can be seen when looking at the electron spectrum for the radiative-decay case in Fig.~\ref{fig:comp_radAuger} (blue dots and line). Electrons occur in this range of energies only due to the deceleration effect of post-collision interaction, as the corresponding spectrum (Fig.~\ref{fig:comp_radAuger}, yellow dots and line) indicates. As mentioned above, the amount of deceleration of the photoelectron due to the interaction with the Auger electron is related to the exact instant of the decay. This can be easily understood in a simplified picture of the PCI process. As the Auger electron is emitted, the charge of the parent ion changes from +1 to +2, which means that the potential $V$ from which the photoelectron emerges switches from $V\sim1/r$ at first to $V\sim2/r$ after the decay. Accordingly, the energy that the photoelectron looses due to this change of the potential depends on the distance the photoelectron has traveled away from the parent ion after its ejection and prior to the Auger-Meitner decay: for a short travel distance the energy loss is large, and if the decay occurs after longer times, i.e., at farther travel distances the energy loss is small. This simplified picture of the PCI process shows that the decay time of the individual event rather than the mean lifetime of the state is encoded in the amount of deceleration of the photoelectron \cite{Schuette2012, Guillemin2012}. We have exploited this fact in previous works to record, for example, the temporal evolution of systems undergoing interatomic Coulombic decay \cite{Trinter2013, Trinter2022}. In the present Letter, it shows that the actual lifetime of the K-shell-ionized cation depends on the kinetic energy release, and thus on the distinct decay route of the system, and in more detail on the internuclear distance of the C and O atoms at the instant of the decay. In order to obtain a more quantitative result, we employed our fitting procedure and fitted PCI line profiles \cite{Armen:1987} for each value of the KER to the measured photoelectron spectrum.

\begin{figure*}[htbp]
    \includegraphics[width=1.0\linewidth]{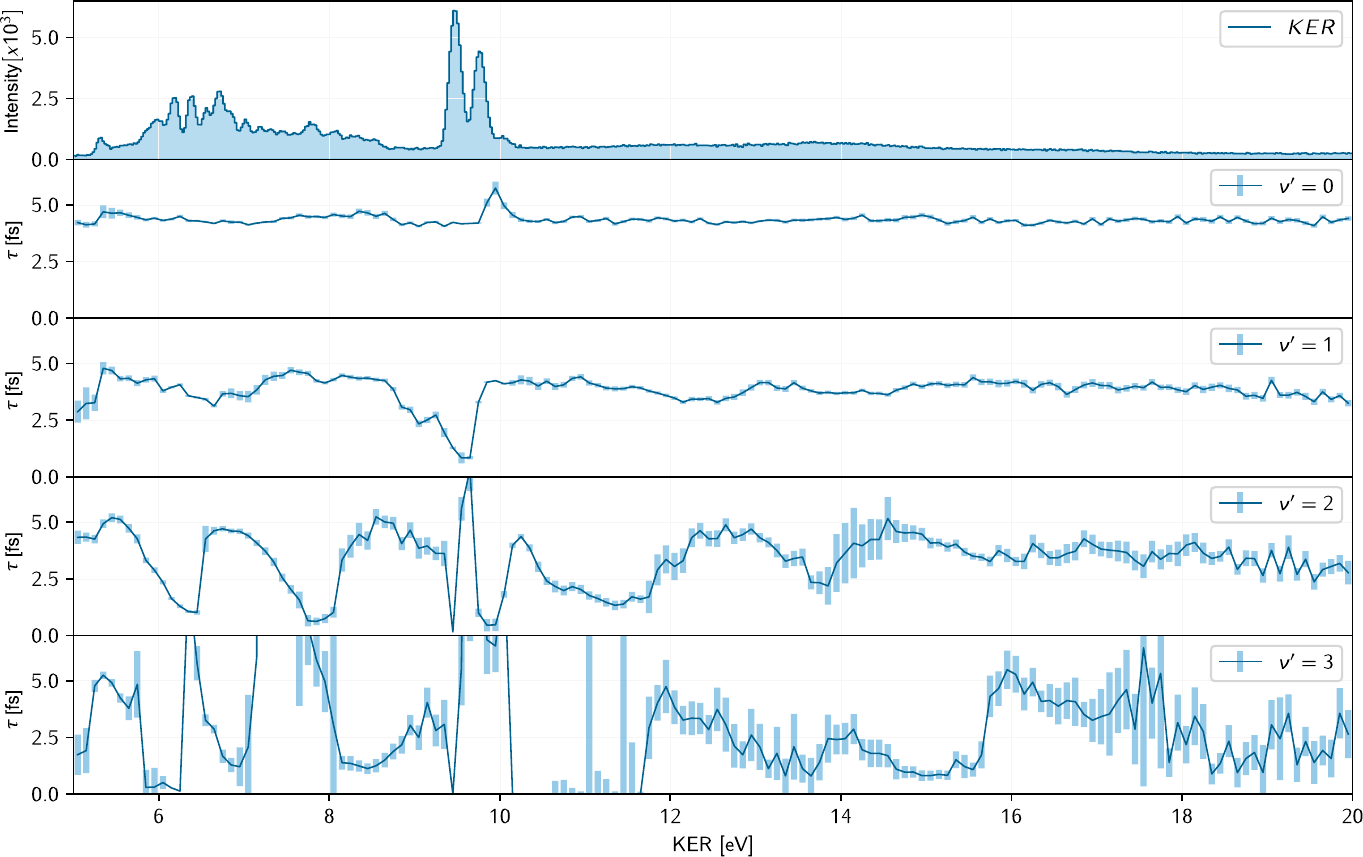}
    \caption{\label{fig:decaytime_ker}Vibrationally resolved lifetimes of the CO C~1s hole in dependence on the KER of the fragment ions. The decay times are obtained from fitting of line profiles \cite{Armen:1987} in the photoelectron-energy spectra, please see text for details. Top to bottom: KER distribution, KER-dependent lifetimes $\tau_n$ for the vibrational states $\nu' = 0...3$.}
\end{figure*}

In the procedure, we used the intensity parameters $a_n$ and $\tau_{\nu' = 0...4}$ as free parameters, which allowed us to extract the lifetimes $\tau_n$ for each vibrational state of the cation separately. Figure~\ref{fig:decaytime_ker} shows the corresponding results. The top panel depicts the KER distribution as a reference, the panels below show $\tau_n$ in dependence of the KER for the four vibrational states $\nu' = 0...3$. The error bars indicate the standard deviation reported by the fitting procedure. The lifetime of the lowest vibrational state $\nu' = 0$ is basically independent from the KER of the molecular fragment ions. The only visible deviation around 10~eV KER has a comparatively high standard deviation and can be explained by the disappearing contribution of the vibrational ground state to the overall spectrum at this point, which leads to problems with the fitting. In contrast, the lifetime changes significantly depending on the KER for the higher-lying vibrational levels of the cation.

\begin{table}[h!]
\begin{tabular}{ |c|c|c|c| } 
 \hline
 Internuclear distance & 1.06~\AA & 1.13~\AA & 1.16~\AA \\
 \hline
 Line width [meV] & 108 & 103 & 98 \\ 
 Lifetime [fs] & 6.10 & 6.40 & 6.72 \\ 
 \hline
\end{tabular}
 \caption{\label{tab:lifetimes}Calculated lifetimes of C~1s$^{-1}$ decaying states for different internuclear distances between the C and O atoms of the CO molecule.}
\end{table}

The results shown in Fig.~\ref{fig:decaytime_ker} demonstrate in general a strong dependence of the lifetime of the K-shell vacancy in the cationic state on the exact decay pathway. For the molecule as a whole this establishes a non-exponential decay behavior. The lifetime of an Auger-Meitner-decaying state depends on the overlap of the orbitals contributing to the decay. Accordingly, it can be expected that the lifetime varies with internuclear distances. The Auger-Meitner decay widths as a function of the internuclear distance were computed using the Fano-CI-Stieltjes approach (see Ref.~\cite{Miteva2017} for details). We employed an aug-cc-pvtz basis set, augmented with 7 s, p, d even tempered basis functions. Furthermore, the molecular orbitals were obtained from a restricted open-shell Hartree-Fock method where we forced an electron to be removed from the 1s orbital of C. The results are shown in Table~\ref{tab:lifetimes}. Our calculations find a decrease of the lifetime with decreasing internuclear distance. This allows for an intuitive interpretation of some of the experimentally observed features, when inspecting the involved potential energy curves shown in Fig.~\ref{fig:Pot_curve}.

For example, there are several features in Fig.~\ref{fig:KER_Ee_histo}(b), which have a \textit{triangular} (or \textit{diagonal}) shape, i.e., features where the measured photoelectron energy decreases as the KER increases. In Fig.~\ref{fig:Pot_curve}, we show as an inset a subset of Fig.~\ref{fig:KER_Ee_histo}(b), which we binned more coarsely and smoothed in order to visually highlight these areas (and to reduce the visibility of the individual vibrational levels in the photoelectron spectrum). Corresponding examples for such triangular features are labeled as \textcircled{1}, \textcircled{2}, and \textcircled{3} in the inset. The KER region belonging to \textcircled{1} is populated as the cation decays initially within in the Franck-Condon region onto the dark blue (a$^1\Sigma^+$) and yellow (b$^1\Pi$) curves in Fig.~\ref{fig:Pot_curve} (labeled as \textcircled{1}, as well), which then finally couple to the C$^+$~+~O$^+$ ground state \cite{Cederbaum1991, Lundqvist1995, Schimmelpfennig1996, Weber2003}. The higher KER in the region connected to feature \textcircled{2} results from the decay onto a set of curves of which the lowest one (c$^1\Delta$) is plotted in Fig.~\ref{fig:Pot_curve} (labeled there as \textcircled{2}, as well). The feature \textcircled{3} corresponds to a decay to a higher-lying potential energy curve. All three KER regions have in common that the potential energy curves are repulsive within the Franck-Condon region of the decay. Accordingly, the measured KER contains information on the internuclear distance | if the repulsive curves are populated at shorter internuclear distances (within the Franck-Condon region) a larger amount of potential energy is converted to KER than in cases where the transition occurs at larger internuclear distances. This behavior generates the aforementioned \textit{triangular/diagonal} features in Fig.~\ref{fig:KER_Ee_histo}(b) (and Fig.~\ref{fig:Pot_curve}, inset). The overall dependence of the lifetime of the internuclear distance indicated in Table~\ref{tab:lifetimes} suggests that at shorter internuclear distances (larger KERs) the lifetime is shorter, and thus the electrons are more strongly decelerated towards lower energies by the PCI. The opposite is true for larger internuclear distances (smaller KERs), where the electrons are less decelerated due to PCI because of the longer lifetime of the decaying state. While the shape of the distribution shown in Fig.~\ref{fig:KER_Ee_histo}(b) is affected, as well, (of course) by the population (and decay) probability of the different vibrational states as such, the aforementioned relation between the internuclear distance, the corresponding KER, and the lifetime can be seen in the extracted $\tau_n$ in Fig.~\ref{fig:decaytime_ker} to some extent directly, as well. Finally, the lack of lifetime variation of the vibrational ground state in Fig.~\ref{fig:decaytime_ker} results from the comparably narrow distribution of internuclear distances. Accordingly, the lack of different internuclear distances contributing to the decay yields an almost constant lifetime of the vibrational ground state of the cation.

In conclusion, we have demonstrated by means of a coincident measurement of the photoelectron and kinetic energy release spectrum that the Auger-Meitner decay of a simple molecule is not of an exponential nature. While this fact is typically neglected and the decay process is modeled in most cases as an exponential one, the coincidence measurement clearly shows the corresponding deviations. These stem from the dependence of the Auger-Meitner decay rate on the internuclear distance of the atoms of the molecule. We have extracted individual lifetimes, which cover a range down to as low as 1~fs. In some cases, the change of the lifetime with the kinetic energy release can be attributed directly to the shape of the potential energy curves involved in the individual decay routes.

\begin{acknowledgements}
We thank the Helmholtz-Zentrum Berlin für Materialien und Energie for the allocation of synchrotron radiation beamtime. We would like to thank the staff of the BESSY~II synchrotron for their excellent support throughout many years and during many beamtimes, in particular the staff of beamline U49-2\_PGM-1, Ronny Golnak and Jie Xiao. This work was supported by the Deutsche Forschungsgemeinschaft. F.T. acknowledges funding by the Deutsche Forschungsgemeinschaft (DFG, German Research Foundation) - Project 509471550, Emmy Noether Programme. T.J. would like to thank Lorenz Cederbaum for helpful discussions on the details of the Auger-Meitner decay and many other topics.
\end{acknowledgements}


\end{document}